\documentclass[prl,twocolumn,superscriptaddress,showpacs,amssymb,amsmath,amsfonts,aps]{revtex4} 
\usepackage[dvips]{color}
\usepackage{graphicx}
\usepackage{dcolumn}

\newcommand{\xbj}{x}

\begin{document}

\title{{\large Measurement of Single and Double  Spin Asymmetries in \\
Deep Inelastic Pion Electroproduction with a Longitudinally Polarized Target}}

 
\newcommand*{\JLAB}{Thomas Jefferson National Accelerator Facility, Newport News, Virginia 23606}
\newcommand*{\JLABindex}{1}
\affiliation{\JLAB}
\newcommand*{\ANL}{Argonne National Laboratory, Argonne, Illinois 60441}
\newcommand*{\ANLindex}{2}
\affiliation{\ANL}
\newcommand*{\ASU}{Arizona State University, Tempe, Arizona 85287-1504}
\newcommand*{\ASUindex}{3}
\affiliation{\ASU}
\newcommand*{\CSUDH}{California State University, Dominguez Hills, Carson, CA 90747}
\newcommand*{\CSUDHindex}{4}
\affiliation{\CSUDH}
\newcommand*{\CANISIUS}{Canisius College, Buffalo, NY}
\newcommand*{\CANISIUSindex}{5}
\affiliation{\CANISIUS}
\newcommand*{\CMU}{Carnegie Mellon University, Pittsburgh, Pennsylvania 15213}
\newcommand*{\CMUindex}{6}
\affiliation{\CMU}
\newcommand*{\CUA}{Catholic University of America, Washington, D.C. 20064}
\newcommand*{\CUAindex}{7}
\affiliation{\CUA}
\newcommand*{\SACLAY}{CEA, Centre de Saclay, Irfu/Service de Physique Nucl\'eaire, 91191 Gif-sur-Yvette, France}
\newcommand*{\SACLAYindex}{8}
\affiliation{\SACLAY}
\newcommand*{\CNU}{Christopher Newport University, Newport News, Virginia 23606}
\newcommand*{\CNUindex}{9}
\affiliation{\CNU}
\newcommand*{\UCONN}{University of Connecticut, Storrs, Connecticut 06269}
\newcommand*{\UCONNindex}{10}
\affiliation{\UCONN}
\newcommand*{\EDINBURGH}{Edinburgh University, Edinburgh EH9 3JZ, United Kingdom}
\newcommand*{\EDINBURGHindex}{11}
\affiliation{\EDINBURGH}
\newcommand*{\FU}{Fairfield University, Fairfield CT 06824}
\newcommand*{\FUindex}{12}
\affiliation{\FU}
\newcommand*{\FIU}{Florida International University, Miami, Florida 33199}
\newcommand*{\FIUindex}{13}
\affiliation{\FIU}
\newcommand*{\FSU}{Florida State University, Tallahassee, Florida 32306}
\newcommand*{\FSUindex}{14}
\affiliation{\FSU}
\newcommand*{\GWU}{The George Washington University, Washington, DC 20052}
\newcommand*{\GWUindex}{15}
\affiliation{\GWU}
\newcommand*{\ISU}{Idaho State University, Pocatello, Idaho 83209}
\newcommand*{\ISUindex}{16}
\affiliation{\ISU}
\newcommand*{\INFNFR}{INFN, Laboratori Nazionali di Frascati, 00044 Frascati, Italy}
\newcommand*{\INFNFRindex}{17}
\affiliation{\INFNFR}
\newcommand*{\INFNGE}{INFN, Sezione di Genova, 16146 Genova, Italy}
\newcommand*{\INFNGEindex}{18}
\affiliation{\INFNGE}
\newcommand*{\INFNRO}{INFN, Sezione di Roma Tor Vergata, 00133 Rome, Italy}
\newcommand*{\INFNROindex}{19}
\affiliation{\INFNRO}
\newcommand*{\ORSAY}{Institut de Physique Nucl\'eaire ORSAY, Orsay, France}
\newcommand*{\ORSAYindex}{20}
\affiliation{\ORSAY}
\newcommand*{\ITEP}{Institute of Theoretical and Experimental Physics, Moscow, 117259, Russia}
\newcommand*{\ITEPindex}{21}
\affiliation{\ITEP}
\newcommand*{\JMU}{James Madison University, Harrisonburg, Virginia 22807}
\newcommand*{\JMUindex}{22}
\affiliation{\JMU}
\newcommand*{\KNU}{Kyungpook National University, Daegu 702-701, Republic of Korea}
\newcommand*{\KNUindex}{23}
\affiliation{\KNU}
\newcommand*{\LPSC}{LPSC, Universite Joseph Fourier, CNRS/IN2P3, INPG, Grenoble, France
}
\newcommand*{\LPSCindex}{24}
\affiliation{\LPSC}
\newcommand*{\MIT}{Massachusetts Institute of Technology, Cambridge, Massachusetts  02139-4307}
\newcommand*{\MITindex}{25}
\affiliation{\MIT}
\newcommand*{\UNH}{University of New Hampshire, Durham, New Hampshire 03824-3568}
\newcommand*{\UNHindex}{26}
\affiliation{\UNH}
\newcommand*{\NSU}{Norfolk State University, Norfolk, Virginia 23504}
\newcommand*{\NSUindex}{27}
\affiliation{\NSU}
\newcommand*{\OHIOU}{Ohio University, Athens, Ohio  45701}
\newcommand*{\OHIOUindex}{28}
\affiliation{\OHIOU}
\newcommand*{\ODU}{Old Dominion University, Norfolk, Virginia 23529}
\newcommand*{\ODUindex}{29}
\affiliation{\ODU}
\newcommand*{\RPI}{Rensselaer Polytechnic Institute, Troy, New York 12180-3590}
\newcommand*{\RPIindex}{30}
\affiliation{\RPI}
\newcommand*{\URICH}{University of Richmond, Richmond, Virginia 23173}
\newcommand*{\URICHindex}{31}
\affiliation{\URICH}
\newcommand*{\ROMAII}{Universita' di Roma Tor Vergata, 00133 Rome Italy}
\newcommand*{\ROMAIIindex}{32}
\affiliation{\ROMAII}
\newcommand*{\MSU}{Skobeltsyn Nuclear Physics Institute,  119899 Moscow, Russia}
\newcommand*{\MSUindex}{33}
\affiliation{\MSU}
\newcommand*{\SCAROLINA}{University of South Carolina, Columbia, South Carolina 29208}
\newcommand*{\SCAROLINAindex}{34}
\affiliation{\SCAROLINA}
\newcommand*{\UNIONC}{Union College, Schenectady, NY 12308}
\newcommand*{\UNIONCindex}{35}
\affiliation{\UNIONC}
\newcommand*{\UTFSM}{Universidad T\'{e}cnica Federico Santa Mar\'{i}a, Casilla 110-V Valpara\'{i}so, Chile}
\newcommand*{\UTFSMindex}{36}
\affiliation{\UTFSM}
\newcommand*{\GLASGOW}{University of Glasgow, Glasgow G12 8QQ, United Kingdom}
\newcommand*{\GLASGOWindex}{37}
\affiliation{\GLASGOW}
\newcommand*{\VIRGINIA}{University of Virginia, Charlottesville, Virginia 22901}
\newcommand*{\VIRGINIAindex}{38}
\affiliation{\VIRGINIA}
\newcommand*{\WM}{College of William and Mary, Williamsburg, Virginia 23187-8795}
\newcommand*{\WMindex}{39}
\affiliation{\WM}
\newcommand*{\YEREVAN}{Yerevan Physics Institute, 375036 Yerevan, Armenia}
\newcommand*{\YEREVANindex}{40}
\affiliation{\YEREVAN}

\newcommand*{\NOWCUA}{Catholic University of America, Washington, D.C. 20064}
\newcommand*{\NOWUK}{University of Kentucky, Lexington, Kentucky 40506}
\newcommand*{\NOWJLAB}{Thomas Jefferson National Accelerator Facility, Newport News, Virginia 23606}
\newcommand*{\NOWUTFSM}{Universidad T\'{e}cnica Federico Santa Mar\'{i}a, Casilla 110-V Valpara\'{i}so, Chile}
\newcommand*{\NOWLANL}{Los Alamos National Laborotory, New Mexico, NM}
\newcommand*{\NOWGWU}{The George Washington University, Washington, DC 20052}
\newcommand*{\NOWCNU}{Christopher Newport University, Newport News, Virginia 23606}
\newcommand*{\NOWORSAY}{Institut de Physique Nucl\'eaire ORSAY, Orsay, France}
\newcommand*{\NOWEDINBURGH}{Edinburgh University, Edinburgh EH9 3JZ, United Kingdom}
\newcommand*{\NOWWM}{College of William and Mary, Williamsburg, Virginia 23187-8795}

  
\author{H.~Avakian}
\affiliation{\JLAB}
\author{P.~Bosted}
\affiliation{\JLAB}
\author{V.D.~Burkert}
\affiliation{\JLAB}
\author{L.~Elouadrhiri}
\affiliation{\JLAB}
\author {K. P. ~Adhikari} 
\affiliation{\ODU}
\author {M.~Aghasyan} 
\affiliation{\INFNFR}
\author {M.~Amaryan} 
\affiliation{\ODU}
\author {M.~Anghinolfi} 
\affiliation{\INFNGE}
\author {H.~Baghdasaryan} 
\affiliation{\VIRGINIA}
\author {J.~Ball} 
\affiliation{\SACLAY}
% ...

\collaboration{The CLAS Collaboration}
\noaffiliation

\date{\today}
\begin{abstract}
We report the first measurement of the transverse momentum dependence of double spin 
asymmetries in semi-inclusive production of pions in deep inelastic scattering
off the longitudinally polarized proton.
Data have been obtained using a polarized electron beam of 5.7 GeV
with the CLAS detector at the Thomas Jefferson National Accelerator
Facility (JLab). 
A significant non-zero $\sin2\phi$ single spin asymmetry was also observed for the first time
indicating strong spin-orbit correlations for transversely polarized quarks in the
longitudinally polarized proton.
The azimuthal modulations of single spin asymmetries have been measured
over a wide kinematic range.
\end{abstract}
\maketitle

A measurement of transverse momenta ($P_T$) of final-state 
hadrons in semi-inclusive deep inelastic scattering
(SIDIS) $\vec{e} \vec{p} \rightarrow e^\prime h X$, for which a hadron is detected in coincidence with the scattered 
lepton, gives access to the 
transverse momentum distributions (TMDs) of partons, which are not accessible 
in inclusive scattering. 
QCD factorization for SIDIS, established at low transverse momentum in the current-fragmentation 
region at higher energies~\cite{Collins:1981uk,Ji:2004wu,Collins:2004nx}, provides a rigorous 
starting point for the study of partonic 
TMDs from SIDIS data using different spin-dependent and spin-independent observables~\cite{Bacchetta:2006tn}.

Measurements of the $P_T$-dependences of spin asymmetries (for $P_T$ comparable to the proton mass $M_p$ and $\Lambda_{QCD}$), in particular, 
allow studies of  transverse momentum ($k_T$) widths of different TMDs, providing  quantitative information on how quarks are 
confined in hadrons.

The $P_T$-dependence of the double-spin asymmetry
also probes the transition from a non-perturbative to a perturbative description.
At large $P_T$ ($\Lambda_{QCD}<<P_T<<Q$), the double spin asymmetry  
is expected to be independent of $P_T$ \cite{Ji:2004wu}. 

Azimuthal distributions of final state particles 
in SIDIS are sensitive to the orbital 
motion of quarks and 
play an important role in the study of transverse momentum distributions of 
quarks in the nucleon.
Large Single Spin Asymmetries (SSAs), appearing as azimuthal moments of the 
cross section, have been observed for decades in hadronic reactions. They  have been among 
the most difficult phenomena to understand from first principles in QCD.
Two fundamental mechanisms have been identified that lead to SSAs in hard processes;
the Sivers mechanism \cite{Sivers:1990fh,Anselmino:1998yz,Brodsky:2002cx,Collins:2002kn,Ji:2002aa}, which generates 
an asymmetry in the distribution of quarks due to orbital motion
of partons, and the Collins
mechanism \cite{Collins:2002kn,Mulders:1995dh}, which generates 
an asymmetry  during the hadronization of quarks.

Measurements of significant azimuthal asymmetries have been reported 
for pion production  in semi-inclusive deep-inelastic scattering 
by the HERMES and COMPASS Collaborations, as well as the CLAS and Hall-C Collaborations at JLab 
for different combinations of beam and target polarizations
\cite{Airapetian:1999tv,Airapetian:2001eg,Airapetian:2004tw,Airapetian:2006rx,Giordano:2009hi,Alexakhin:2005iw,Kafer:2008ud,
Avakian:2003pk,Avakian:2005ps,Mkrtchyan:2007sr,Osipenko:2008rv}.

For the longitudinally polarized target case, first discussed by Kotzinian and 
Mulders \cite{Mulders:1995dh,Kotzinian:1994dv,Kotzinian:1995cz}, the only SSA,
depending on the azimuthal angle $\phi$ between the lepton scattering and pion production 
planes \cite{Bacchetta:2004jz},
arising at leading order is the $\sin2\phi$ moment.
For a given Bjorken variable ($\xbj$) and  fraction of the energy of 
the virtual photon carried by the final state hadron ($z$), it involves the convolution of
distribution and fragmentation functions. Corresponding functions are  
the Ralston-Soper-Mulders-Tangerman (RSMT) distribution function $h_{1L}^\perp(x,k_T)$ \cite{Ralston:1979ys,Mulders:1995dh} describing
the transverse polarization of quarks in a longitudinally polarized proton 
\cite{Mulders:1995dh,Kotzinian:1994dv,Kotzinian:1995cz,Ji:2004wu,DiSalvo:2006bq}, and the
Collins fragmentation function $H_1^\perp(z,p_T)$
 \cite{Collins:1992kk} describing fragmentation of transversely polarized quarks into 
unpolarized hadrons. The final transverse momentum of the hadron in leading order is 
defined by the combination $zk_T+p_T$, where $p_T$ is the transverse momentum generated in the
hadronization process. 



The only available measurement of the $\sin 2\phi$ moment  by
  HERMES \cite{Airapetian:1999tv} is consistent with zero. 
The RSMT distribution function  has been studied in various
QCD inspired models \cite{Gamberg:2007wm,Avakian:2007mv,Efremov:2009ze,Boffi:2009sh}. First calculations
for $h_{1L}^\perp(x,k_T)$ have recently been performed in the perturbative limit  \cite{Zhou:2009jm},
and first measurements   have been performed using lattice methods
\cite{Hagler:2009mb}.  
A measurably large asymmetry has been predicted~\cite{Efremov:2002ut,Gamberg:2007wm,Avakian:2007mv,Efremov:2009ze,Boffi:2009sh} only at large $x$ ($x>0.2$), a region well-covered by JLab. 
The same distribution function is also accessible in double-polarized Drell-Yan production,
where it gives rise to the $\cos2\phi$ azimuthal moment 
in the cross section \cite{Tangerman:1994eh}.

The $\sin \phi$   moment of the spin-dependent cross section
for the longitudinally polarized target  is dominated by
higher-twist contributions \cite{Bacchetta:2006tn} which are suppressed by $1/Q$ at 
large momentum transfer.  This moment has been measured for the first time by the 
HERMES Collaboration \cite{Airapetian:1999tv}.
Higher-twist observables, such as longitudinally polarized beam or target SSAs, 
are important  for understanding long-range quark-gluon dynamics. 
Recently, higher-twist effects in SIDIS were interpreted in terms of 
an average transverse force acting on the active quarks in  the instant after
being struck by the virtual photon \cite{Burkardt:2008vd}.

Both $\sin\phi$ and $\sin 2\phi$ moments of the SIDIS cross section for 
longitudinally polarized targets can be an
important source of independent information on the Collins fragmentation 
mechanism~\cite{Bacchetta:2006tn}, complementary to recent Belle  measurements~\cite{Abe:2005zx}. The  
$\sin 2\phi$ asymmetry, however, provides a 
cleaner measurement of Collins fragmentation because it
doesn't have a Sivers type contribution in the leading order~\cite{Kotzinian:1995cz}.

In this Letter, we present measurements of the kinematic dependences of
different single- and double-spin asymmetries in semi-inclusive pion production
off longitudinally polarized protons.
The current analysis is based on recently published data~\cite{Dharmawardane:2006zd} 
from Jefferson Lab. The CEBAF Large Acceptance 
Spectrometer~\cite{Mecking:2003zu} in Jefferson Lab's Hall B 
was used to measure spin asymmetries in the scattering of 
longitudinally polarized electrons from 
longitudinally polarized protons.
The data were collected in  2001 using an incident beam of 5-nA  with $E=5.7$ GeV energy
and an average beam polarization of $P_B=70$\%.
The detector package~\cite{Mecking:2003zu}
provided a clean identification of electrons scattered at polar
angles between 8 and 45 degrees. 
Charged and neutral pions were identified using the  
time-of-flight from  the target to the timing scintillators and 
the signal in the lead-scintillator electromagnetic calorimeter, respectively.
Ammonia ($^{15}$NH$_3$), polarized via 
Dynamic Nuclear Polarization~\cite{Keith:2003ca}, was used
to provide polarized protons.
The average target polarization ($P_t$) was about 75\%. 
The data were divided
into 5 bins in $Q^2$ (0.9 - 5.4 GeV$^2)$, 6 bins in $x$ (0.12 - 0.48), 
3 bins in $z$ (0.4 - 0.7), 9 bins in $P_T$ (0 - 1.12 GeV/c) and 12 bins in $\phi$ (0 - 2$\pi$).
Cuts on the missing mass of $e'\pi X$ ($M_X>1.4$ GeV) and on the fraction of the virtual
photon  energy $\nu$ carried by the pion $z$ ($z<0.7$), have been used to suppress the contribution from exclusive processes.
At large $z$  ($z>0.7$)  the fraction of $\pi^\pm$ from $\rho^0$-decays can be fairly large  and the 
corrections due to pions coming from the $\rho$ (from 5 to 20\% for $z<0.7$), not accounted for in the current analysis, 
may be significant.

The double spin asymmetry $A_1$ is defined as 
\begin{eqnarray} 
A_1= \frac{1}{f D^{\prime}(y) P_BP_t}\frac{N^{+}-N^{-}}{N^{+}+N^{-}}
\end{eqnarray} 
\noindent
where $f\approx 0.14$  (dependent on kinematics) is the dilution factor,  $y=\nu/E$, 
and $N^{\pm}$ are luminosity-weighted counts for antiparallel and parallel electron and proton helicities. 
The contribution from the longitudinal photon 
 is accounted for in the  depolarization factor  $D^{\prime}(y)$:

\begin{eqnarray} 
 D^{\prime}(y)=\frac{(1-\varepsilon)(2-y)}{y(1+\varepsilon R)} \equiv  \frac{y(2-y)}{y^2+2\left( 1-y-\frac{y^2\gamma^2}{4}\right)\frac{(1+R)}{(1+\gamma^2)}},
\label{FLSIGMC}
\end{eqnarray}
\noindent where $R$~\cite{Dasu:1988ru} is the ratio of longitudinal to transverse 
photon contributions
and $\varepsilon$ is the ratio of longitudinal and transverse photon fluxes.

\begin{figure}[h]
\includegraphics[height=.25\textheight]{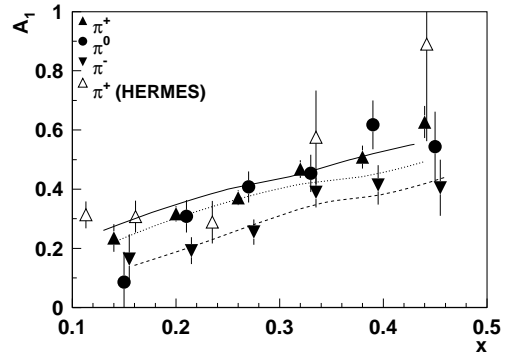}
\caption{\small 
The double-spin asymmetry as a function of $x$ from a polarized proton target
 for different pions. Open triangles correspond to the HERMES measurement of
$A_1$ for $\pi^+$ \protect\cite{Airapetian:2004zf}. Only statistical uncertainties are shown.
The solid, dashed and dotted curves, calculated using
LO GRSV PDF \cite{Gluck:1995yr} and 
$D_1^{d\rightarrow \pi+}/D_1^{u\rightarrow \pi+}=1/(1+z)^2$ \cite{Mkrtchyan:2007sr} correspond to
$\pi^+$, $\pi^-$, and $\pi^0$, respectively.
}
\label{a1p3pionsxq47}
\end{figure}

The main sources of systematic uncertainties in the measurements of the double
spin asymmetries include uncertainties in
beam and target polarizations (4\%), dilution factor (5\%), and
depolarization factor (5\%). Contributions from target fragmentation, 
kaon contamination and radiative corrections~\cite{Akushevich:1997di} were estimated to be below $ 3\%$ each.

The dependence of the double-spin asymmetry on Bjorken $x$ for different pions
obtained from the CLAS data 
is presented in Fig. \ref{a1p3pionsxq47}.
The results for  $A_1$ are consistent with the
HERMES semi-inclusive data, and
at large $x$ have significantly smaller statistical uncertainties.
The double spin asymmetries 
measured  by HERMES and CLAS at different beam energies (by a factor of $\approx 5$)  and different 
values of average $Q^2$  (by a factor of $\approx 3$), for a fixed $x$-bin are in good agreement, indicating no
significant $Q^2$ dependence of the double polarization asymmetry $A_1$.  
Measured asymmetries are also consistent with calculations performed 
 using leading-order GRSV PDFs \cite{Gluck:1995yr} and a simple parametrization
of the ratio of unfavored and favored fragmentation functions~\cite{Mkrtchyan:2007sr}.

\begin{figure}[hbt]
\vspace{-3.0cm}
\includegraphics[width=0.5\textwidth]{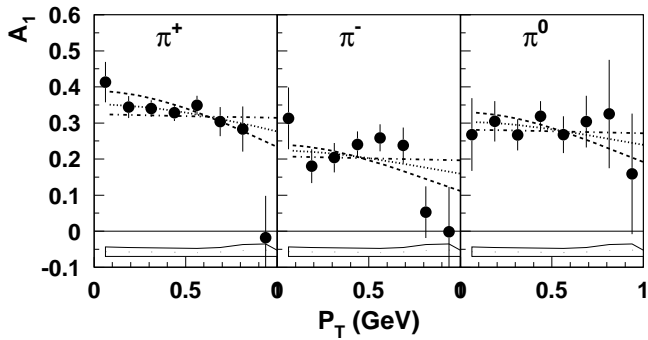}   
\caption{\small 
The double spin asymmetry $A_{1}$ 
as a function of transverse momentum $P_T$,  
integrated over all kinematical variables. The open band corresponds to systematic uncertainties. The dashed, dotted and dash-dotted
curves are calculations for different values for the ratio of transverse momentum widths for $g_1$ and  $f_1$  
(0.40, 0.68, 1.0) for  
a fixed width for $f_1$ (0.25 GeV$^2$)~\cite{Anselmino:2006yc}. 
}
 \label{a1pptdepvszq47x16}
\end{figure}

$A_1$ is shown in Fig.~\ref{a1pptdepvszq47x16} 
as a function of $P_T$, integrated over all $\xbj$ (0.12--0.48) for $Q^2>1$ GeV$^2$,  
$W^2>4$ GeV$^2$, and $y<0.85$.
Although these plots are consistent with flat distributions, $A_1(P_T)$ may decrease somewhat with $P_T$ at moderately small $P_T$ 
for $\pi^+$. The slope for $\pi^-$ could be positive for 
moderate $P_T$ (ignoring the first data point).

A possible interpretation of the $P_T$-dependence of the double-spin asymmetry
may involve different widths of the transverse momentum distributions of quarks
with different flavor and polarizations \cite{Anselmino:2006yc} resulting from different
orbital motion of quarks polarized in the direction of the proton spin and 
opposite to it \cite{Brodsky:1994kg,Avakian:2007xa}. In Fig.~\ref{a1pptdepvszq47x16} the
measured $A_1$ is compared with calculations of the Torino group \cite{Anselmino:2006yc}, 
which uses different values of
the ratio of widths in $k_T$  for partonic helicity, $g_1$, and momentum, $f_1$, distributions,
assuming Gaussian $k_T$ distributions  with no flavor dependence.
A fit to $A_1(P_T)$ for $\pi^+$  using the same approach yields a ratio of widths of  $0.7\pm0.1$
with  $\chi^2=1.5$. The fit to $A_1$ with a straight line (no difference in $g_1$ and $f_1$ widths) gives a $\chi^2=1.9$.



Asymmetries as a function of the azimuthal angle $\phi$ provide access
to different combinations of TMD parton distribution and fragmentation 
functions~\cite{Bacchetta:2006tn}.
The longitudinally polarized (L) target  spin asymmetry for an unpolarized 
beam (U), 
\begin{eqnarray}
A_{UL}= \frac{1}{f P_t}\frac{N^{+}-N^{-}}{N^{+}+N^{-}}
\end{eqnarray}
is measured  from data by counting in $\phi$-bins 
the difference of luminosity-normalized events with proton spin states anti-parallel ($N^+$) and 
parallel ($N^-$) to the beam direction.

The standard procedure for the extraction of the different moments involves sorting   
$A_{UL}$  in bins of $\phi$ and 
fitting this $\phi$-distribution with theoretically motivated functions. Results for the function 
$p_1\sin\phi+p_2\sin 2\phi$
and, alternatively, for $(p_1\sin\phi+p_2\sin 2\phi)/(1+p_3\cos\phi)$ are consistent, 
indicating a weak dependence
of the extracted $\sin n\phi$ moments on the presence of the $\cos \phi$ moment in the $\phi$-dependence of the 
spin-independent sum.
The main sources of systematic uncertainties in the  measurements of single
spin asymmetries include uncertainties in
target polarizations (6\%), acceptance effects (8\%), and  uncertainties in the dilution factor (5\%).
The contribution due to differences between the true luminosity for the two different target spin states is below 2\%.
Radiative corrections for $\sin\phi$-type moments, for moderate values of $y$ are expected to be 
negligible \cite{Akushevich:1999hz}.

\begin{figure}
\vspace{-3.5cm}
\includegraphics[height=.4\textheight,width=.36\textheight]{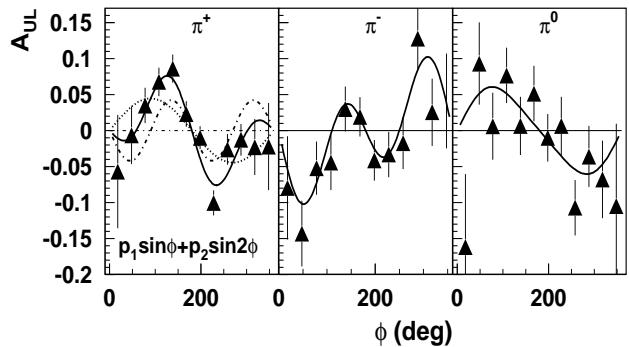}
\caption{Azimuthal modulation of the target single spin asymmetry 
$A_{UL}$ for pions integrated over the full kinematics. Only statistical uncertainties are shown.
Fit parameters $p_1/p_2$ are $0.047\pm0.010/-0.042\pm0.010$,  $-0.046\pm0.016/-0.060\pm0.016$,  $0.059\pm0.018/0.010\pm0.019$ for $\pi^+,\pi^-$ and $\pi^0$, respectively. 
Dotted and dash-dotted lines for $\pi^+$ show separately contributions from $\sin\phi$ and $\sin 2\phi$ 
moments, whereas the solid line shows the sum.
}
\label{fig:sinusoid}
\end{figure}

The dependence of the target single spin asymmetry on $\phi$,
integrated over all other kinematical variables,  is plotted
in Fig.~\ref{fig:sinusoid}.
We observe a significant $\sin 2\phi$ modulation for $\pi^+$ ($0.042\pm0.010$). 
A relatively small $\sin 2\phi$  term in the azimuthal dependence for $\pi^0$ 
is in agreement with observations by HERMES \cite{Airapetian:2004tw}. 
Since the only known contribution to the  $\sin 2\phi$  moments comes from the Collins effect, one can infer that, 
for $\pi^0$, the Collins function is suppressed. Indeed, both HERMES~\cite{Airapetian:2004tw} and Belle~\cite{Abe:2005zx}
 measurements indicate that favored and unfavored Collins functions are roughly equal and have opposite signs, 
which means that they largely cancel for  $\pi^0$. On the other hand, the amplitudes of the   $\sin \phi$ modulations 
for $\pi^+$ and $\pi^0$ are comparable in size.
This indicates that the contribution from the Collins effect to the $\sin \phi$ SSA, in general, is relatively small.

\begin{figure}[htb]
\vspace{-2.0cm}
\begin{center}
\includegraphics[height=.34\textheight]{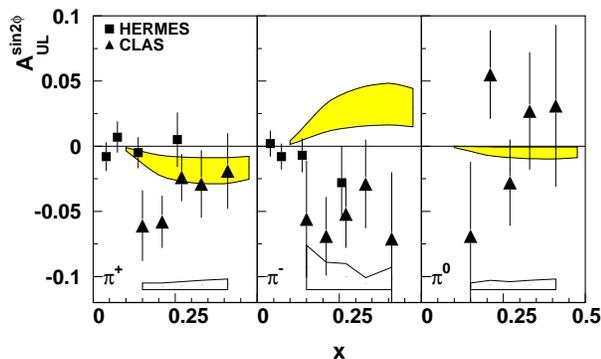}
\end{center}
\caption{The measured $x$-dependence of the longitudinal target SSA  $A_{UL}^{\sin2\phi}$ (triangles).
The squares show the existing measurement of  $A_{UL}^{\sin2\phi}$ from 
HERMES.  The lower band shows the systematic uncertainty. 
The upper band shows the existing theory predictions with uncertainties due to the Collins 
function~\cite{Efremov:2006qm,Avakian:2007mv}.  
\label{fig:aul11.sin2}}
\end{figure}

The  $\sin 2\phi$ moment  ${\it A}^{\sin2\phi}_{UL}$ as a 
function of $\xbj$   is plotted in Fig.~\ref{fig:aul11.sin2}.
Calculations~\cite{Efremov:2002ut,Avakian:2007mv} using $h_{1L}^\perp$ from the chiral quark soliton model 
\cite{Schweitzer:2001sr} and the Collins 
function~\cite{Efremov:2006qm} extracted from HERMES~\cite{Airapetian:2004tw}
and Belle~\cite{Abe:2005zx} data, are plotted as filled bands in Fig.~\ref{fig:aul11.sin2}.
The kinematic dependence of the SSA
for $\pi^+$ from the CLAS data is roughly  
consistent with these predictions. The interpretation of the $\pi^-$ data, which tend
to have SSAs with a sign opposite to expectations, may require accounting for additional contributions (e.g. interference effects 
from exclusive  $\rho^0p$ and $\pi^-\Delta^{++}$ channels).
This will require a detailed study
with higher statistics of both double and single spin  asymmetries from pions 
coming from $\rho$-decays.


The $\sin2\phi$ moment of the $\pi^+$ SSA at large $\xbj$ is 
dominated by $u$-quarks; therefore with
additional input from Belle measurements~\cite{Abe:2005zx} 
on the ratio of unfavored to 
favored Collins fragmentation
functions, it can provide a first glimpse of the twist-2 
TMD function $h_{1L}^\perp$. 

In summary, kinematic dependencies of single and double spin asymmetries
have been measured in a wide kinematic range in $x$ and $P_T$ with CLAS and a longitudinally polarized proton target. 
Measurements of the $P_T$-dependence of  the
double spin asymmetry, performed for the first time, indicate the possibility
of different average transverse momentum for quarks
aligned or anti-aligned with the nucleon spin.
A non-zero $\sin 2\phi$ single-target spin asymmetry is measured for the first time, indicating
that spin-orbit correlations of transversely polarized quarks in the longitudinally 
polarized nucleon may be significant.


%

New, higher statistics measurements of SSAs in SIDIS at CLAS 
\cite{PACeg1} will
allow us to examine the  $Q^2$, $x$, and $P_T$ dependences  of 
azimuthal moments in multi-dimensional bins and investigate the twist nature of different observables. 


We thank A.~Afanasev, S.~Brodsky, A.~Kotzinian, and P. Schweitzer 
for stimulating discussions.
We would like to acknowledge the outstanding efforts of the staff of the 
Accelerator and the Physics Divisions at JLab that made this experiment possible.
This work was supported in part by the U.S. Department of Energy
and the National  Science Foundation, 
the Italian Istituto Nazionale di Fisica Nucleare, the 
 French Centre National de la Recherche Scientifique, 
the French Commissariat \`{a} l'Energie Atomique, 
 and the National Research Foundation of Korea.
The Southeastern Universities Research Association (SURA) operates the 
Thomas Jefferson National Accelerator Facility for the United States 
Department of Energy under contract DE-AC05-06OR23177. 

\bibliography{all_prl}

\end{document}